\documentclass[aps,prl,twocolumn,twoside,superscriptaddress,nofootinbib,preprintnumbers,floatfix]{revtex4-2}

\usepackage[bookmarks=false]{hyperref}
\usepackage{graphicx}
\usepackage{amsmath}
\usepackage{mathrsfs}
\usepackage{xspace}
\usepackage{xcolor}
\usepackage[normalem]{ulem}
\usepackage{comment}
\usepackage{dsfont}
\usepackage{braket}
\usepackage{qcircuit}
\usepackage{amssymb}
\usepackage{float}

\begin{document}


\title{Particle Production, Equilibration, and Quantum Recurrences from Classical Fields}

\author{Iv\'an Cunt\'in}
\email{ivan.cuntin.broullon@usc.es}
\affiliation{Instituto Galego de F\'isica de Altas Enerx\'ias IGFAE, Universidade de Santiago de Compostela, E-15782 Galicia-Spain}

\author{Wenyang Qian}
\email{wqian@ccnu.edu.cn}
\affiliation{Institute of Particle Physics and Key Laboratory of Quark and Lepton Physics (MOE),
Central China Normal University, Wuhan, 430079, Hubei, China}
\affiliation{Instituto Galego de F\'isica de Altas Enerx\'ias IGFAE, Universidade de Santiago de Compostela, E-15782 Galicia-Spain}

\author{Bin Wu}
\email{b.wu@cern.ch}
\affiliation{Instituto Galego de F\'isica de Altas Enerx\'ias IGFAE, Universidade de Santiago de Compostela, E-15782 Galicia-Spain}

\begin{abstract}
We investigate particle production from classical fields, a phenomenon central to the pre-equilibrium dynamics of relativistic heavy-ion collisions and the reheating epoch of the early Universe. Using lattice $\lambda\phi^4$ theory as a proof of principle, we show that this problem is naturally amenable to quantum computation, providing a first-principles framework for nonequilibrium quantum-field dynamics beyond existing approximations. We perform simulations on small spatial lattices, exhausting our available classical computational resources while maintaining a direct mapping to future quantum-computing implementations. We find that particle production is accompanied by equilibration of observables, including the field expectation value, occupation-number distribution, and pressure. The observed equilibration persists for timescales several times longer than the initial equilibration time before the observables resume oscillatory behavior associated with quantum Poincar\'e recurrences. Our results establish a route toward first-principles studies of equilibration in nonequilibrium quantum field theory and provide insight into the search for the smallest possible locally equilibrated quark--gluon systems at hadron colliders.
\end{abstract}


\maketitle

{\it Introduction.}
One of the central challenges in nonequilibrium quantum field theory (QFT) is to understand how equilibration emerges from highly occupied initial states. Owing to their large occupation numbers, such states can often be described approximately by classical bosonic fields and arise in a wide range of physical settings, including the early stages of relativistic heavy-ion collisions at RHIC and the LHC, where the dynamics are dominated by dense gluonic fields~\cite{McLerran:1993ni,McLerran:1993ka,Jalilian-Marian:1996mkd,
Kovchegov:1998bi, Mueller:1999fp}, and the reheating epoch of the early Universe driven by the inflaton field~\cite{Kofman:1994rk,Kofman:1997yn}.

These phenomena are naturally described within the Schwinger--Keldysh formulation of nonequilibrium QFT~\cite{Berges:2004yj,Calzetta:2008iqa}. However, unlike Euclidean lattice QFT, a first-principles treatment of real-time dynamics remains elusive because the oscillatory phase of the Minkowski path integral leads to a severe sign problem~\cite{Alexandru:2020wrj}. Consequently, current studies rely on controlled approximation schemes.

A major challenge is to develop an approximation that remains valid throughout the entire evolution, since the relevant dynamics involve occupation numbers that can change substantially with time. Approaches based on expansions around fixed occupation numbers therefore have a limited range of applicability. The most widely used framework for this problem, the classical-statistical approximation (CSA)~\cite{Tranberg:2003gi,Tranberg:2006ip,Berges:2012us,Berges:2012ev,Tranberg:2012jp,Tranberg:2012qu,Berges:2013eia}, correctly captures the large-occupancy regime but cannot describe the final equilibration toward a genuine quantum thermal state. In particular, classical thermal systems approach the Rayleigh--Jeans distribution, $f_{\rm cl}=T/p$, resulting in a divergent energy density and ultraviolet-cutoff-dependent thermal observables~\cite{Epelbaum:2014mfa}. The leading quantum corrections beyond the classical limit arise from vacuum fluctuations at order $\hbar$~\cite{Mueller:2002gd,Jeon:2004dh,Dusling:2010rm,Epelbaum:2011pc,Epelbaum:2014mfa}, while a systematic treatment beyond this order requires genuine quantum loop effects~\cite{Jackiw:1974cv}.

Attempts to include such $\mathcal{O}(\hbar)$ corrections in CSA simulations have yielded intriguing results, including early pressure isotropization in non-Abelian gauge theories~\cite{Epelbaum:2013ekf}, consistent with phenomenological descriptions of heavy-ion collisions~\cite{Heinz:2013th}. However, these extensions are not renormalizable~\cite{Epelbaum:2014yja} and suffer from ultraviolet divergences~\cite{Berges:2013lsa,Epelbaum:2014mfa}, whose systematic removal remains an open problem.

An alternative framework for addressing these phenomena is the two-particle-irreducible (2PI) effective action formalism~\cite{Berges:2004yj,Calzetta:2008iqa}, which has been highly successful in scalar theories (see Ref.~\cite{Gelis:2024iar} and references therein for recent progress). However, its practical implementations require truncations of the effective action, limiting their applicability in regimes with large field expectation values. Moreover, controlled applications to non-Abelian gauge theories remain challenging because gauge invariance is not preserved at finite truncation order~\cite{Arrizabalaga:2002hn,Berges:2004yj}.

In this work, we propose to overcome these limitations by employing quantum-computing techniques and performing the first quantum simulations of particle production from classical fields in QFT. To reduce the required computational resources, we consider a spin-0 scalar field theory, namely lattice $\lambda\phi^4$ theory. Considerable progress has been made in quantum algorithms and quantum simulations of this theory~\cite{Jordan:2012xnu,Jordan:2011ci,Klco:2018zqz,Barata:2020jtq,Macridin:2021uwn,Li:2022ped,Hardy:2024ric,Ingoldby:2025bdb,Abel:2025pxa,Cao:2026nyj}, culminating in recent hardware demonstrations involving tens to $\mathcal{O}(100)$ qubits~\cite{Zemlevskiy:2024vxt,Farrell:2025nkx}. Lattice scalar field theory has also served as a proof-of-concept application of digital quantum computing to collider physics using Soft-Collinear Effective Theory~\cite{Bauer:2021gup}. Furthermore, quantum simulation offers a route toward lattice gauge theories, as discussed in recent reviews~\cite{Halimeh:2025oeq}. We demonstrate below that quantum computing provides a promising framework for first-principles investigations of these nonequilibrium dynamics relevant to heavy-ion collisions and the reheating epoch of the early Universe.

{\it Setup.—}
We consider lattice $\lambda\phi^4$ theory in $d$ spatial dimensions, with Hamiltonian
\begin{align}
H=
a^d\sum_{\mathbf{x}}
\left[
\frac12\hat\pi_{\mathbf{x}}^2
+\frac12m^2\hat\phi_{\mathbf{x}}^2
+\frac12|\nabla\hat{\phi}_{\mathbf{x}}|^2
+V(\hat{\phi}_{\mathbf{x}})
\right]\,,
\label{eq:H_FB}
\end{align}
where $V(\hat{\phi})\equiv\lambda\hat{\phi}^4/4!$, $a$ is the lattice spacing, and the field operator $\hat{\phi}$ and canonical momentum density $\hat{\pi}$ satisfy
$[\hat\phi_{\mathbf{x}},\hat\pi_{\mathbf{y}}]
=i a^{-d}\delta_{\mathbf{x},\mathbf{y}}\,$.
Introducing lattice creation and annihilation operators~\cite{Jordan:2011ci, Klco:2018zqz, Barata:2020jtq}, the Hamiltonian can be written as
\begin{align}
\label{eq:H_HO}
H
=
\sum_{\mathbf p}\omega_{\mathbf p}
\left(\hat a^\dagger_{\mathbf p}\hat a_{\mathbf p}
+\frac12\right)
+
a^d\sum_{\mathbf x}V(\hat{\phi}_{\mathbf x})\,,
\end{align}
with
\begin{align}
\omega_{\mathbf p}
=
\sqrt{
m^2+\frac{4}{a^2}
\sum_{j=1}^{d}\sin^2\left(\frac{p_j a}{2}\right)
}\,.
\end{align}
Here, periodic boundary conditions are imposed, resulting in discrete lattice momentum modes $\mathbf{p}$.

The highly occupied initial states considered here are naturally represented by coherent states~\cite{Glauber:1963tx}:
\begin{align}
\label{eq:psi0}
|\psi(0)\rangle
=
\bigotimes_{\mathbf p}|\alpha_{\mathbf p}\rangle
~~~~\text{with}~~~~
|\alpha_{\mathbf p}\rangle
=
e^{
\alpha_{\mathbf p}\hat a_{\mathbf p}^{\dagger}
-\alpha_{\mathbf p}^*\hat a_{\mathbf p}}|0\rangle\,,
\end{align}
where $|0\rangle$ is the vacuum state of the free Hamiltonian and $\alpha_{\mathbf p}$ are complex coherent-state parameters. Such states minimize quantum fluctuations and provide a direct mapping to classical configurations~\cite{Dusling:2010rm, Epelbaum:2011pc}:
\begin{align}
\label{eq:phi_pi_cl}
\langle\psi(0)|\hat\phi|\psi(0)\rangle
&=\phi_{\rm cl}(0)\,,
&
\langle\psi(0)|\hat\pi|\psi(0)\rangle
&=\pi_{\rm cl}(0)\,.
\end{align}
These relations determine the coherent-state parameters $\alpha_{\mathbf p}$. In this work, we focus on spatially homogeneous initial configurations, for which only the zero-momentum mode is populated,
$\alpha_{\mathbf p}\propto\delta_{\mathbf p,\mathbf 0}$.

{\it Efficient quantum simulation.—}
Particle production from classical fields admits a polynomially scalable implementation on future fault-tolerant gate-based quantum computers, as detailed in the Supplemental Material. The algorithm consists of three steps: coherent-state preparation, real-time evolution, and measurement of observables. The free vacuum can be prepared using the Kitaev--Webb algorithm~\cite{Kitaev:2008vci}, with polynomial complexity in the lattice size $N$ and the number of qubits per site $n_Q$~\cite{Jordan:2011ci,Bagherimehrab:2021xlp}. Coherent states are then prepared by applying displacement operators to the vacuum. Real-time evolution is implemented using product-formula decompositions of the evolution operator,
$U(t)=e^{-iHt}$,
with the Hamiltonian expressed in the field-operator basis of Eq.~\eqref{eq:H_FB}~\cite{Jordan:2012xnu,Jordan:2011ci,Klco:2018zqz,Li:2022ped}.

For a $p$th-order product formula, the CNOT-gate complexity scales as
\begin{align}
N_{\rm gate}
=
\mathcal{O}\!\left(
N n_Q^4
\frac{t^{1+1/p}}{\epsilon^{1/p}}
\right)\,,
\end{align}
where $\epsilon$ is the target error. The linear dependence on the number of lattice sites contrasts with the exponential cost of exact classical simulations arising from the exponential growth of the many-body Hilbert space. Measurements of local observables, including field expectation values, occupation numbers, and the pressure, also require only polynomial resources.

Current quantum devices remain in the noisy intermediate-scale quantum (NISQ) era~\cite{Preskill:2018jim}. We therefore benchmark the algorithm by directly exponentiating the Hamiltonian to obtain the exact evolution operator, together with ideal quantum-circuit simulations. In the quantum-circuit simulations presented here (see Supplemental Material), which are restricted to small lattices, the initial states are prepared using a variational quantum eigensolver (VQE)~\cite{Peruzzo:2014,McClean:2016} combined with adiabatic state preparation, rather than the Kitaev--Webb algorithm, which is designed for scalable fault-tolerant implementations. These benchmarks establish a direct connection to future fault-tolerant quantum computing and lay the groundwork for first-principles investigations of particle-production dynamics beyond existing approximation schemes, such as the CSA and the 2PI formalism.

In our quantum simulations, expectation values are estimated from repeated
measurements of the quantum state. Field operators are diagonal in the
field-amplitude basis, whereas momentum-space observables are obtained after
applying the quantum Fourier transform. Statistical uncertainties arising from
the finite number of shots are estimated from these measurements, as detailed
in the Supplemental Material.
 The time-dependent field expectation value is
defined as
\begin{align}
\phi_{\rm cl}(t)
=
\langle\psi(t)|\hat{\phi}|\psi(t)\rangle \,,
\end{align}
with
$|\psi(t)\rangle=U(t)|\psi(0)\rangle$. The occupation-number distribution and pressure are
defined as
\begin{align}
f_{\mathbf p}
&=
\langle\psi(t)|
\hat a^\dagger_{\mathbf p}\hat a_{\mathbf p}
|\psi(t)\rangle \,,
\\
P
&=
\big\langle\psi(t)\big|
\big[\frac{\hat{\pi}^2}{2}
+\frac{2-d}{2d}
|\nabla\hat{\phi}|^2
-\frac{m^2}{2}\hat{\phi}^2
-V(\hat{\phi})\big]
\big|\psi(t)\big\rangle.
\notag
\end{align}

{\it Equilibration and quantum recurrences in small systems.—}
We begin by simulating the quantum circuits constructed with \texttt{Qiskit}~\cite{QiskitCommunity2017} and benchmarking them against exact calculations obtained by direct exponentiation of the Hamiltonian in the harmonic-oscillator basis of Eq.~\eqref{eq:H_HO}, for the simplest nontrivial case: a one-dimensional lattice with two spatial sites, corresponding to two momentum modes, $p_0=0$ and $p_1=\pi/a$. The circuits are
executed on an ideal quantum simulator implementing the gate-based quantum
algorithm described above.

\begin{figure}[ht]
\includegraphics[width=0.48\textwidth]{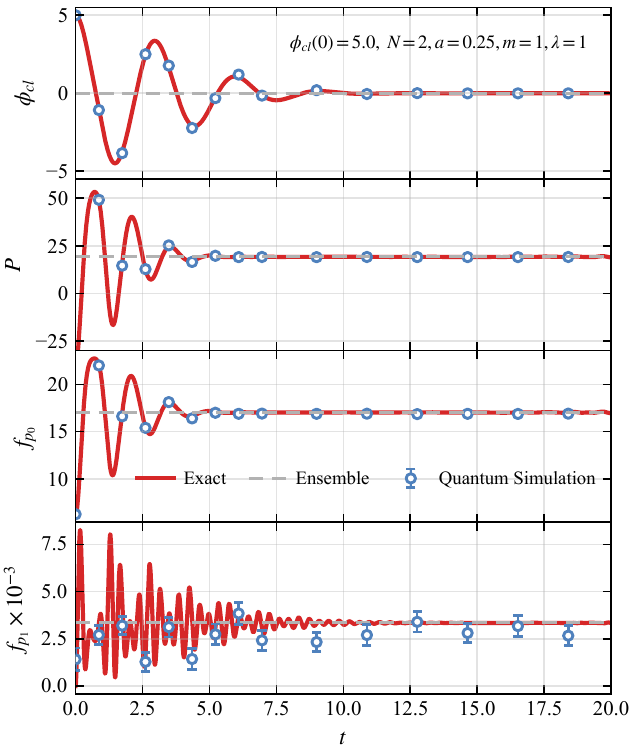}
\caption{
Particle production and equilibration from classical fields. For
$\phi_{\rm cl}(0)=5$, $\pi_{\rm cl}(0)=0$, $a=0.25$, $m=1$ and $\lambda=1$, the field
expectation value $\phi_{\rm cl}$, pressure $P$, and occupation-number
distribution $f_{\mathbf p}$ obtained by direct exponentiation (Exact) all
relax to the values predicted by the diagonal ensemble $\rho_{\rm eq}$
[Eq.~\eqref{eq:rho_eq}, Ensemble]. Data points with error bars (Quantum
Simulation) are obtained from $2^{20}$ measurements of the \texttt{Qiskit}
circuit with $n_Q=6$ qubits per site and second-order Trotter step
$\delta t=0.01$.
}
\label{fig:Phi05N2}
\end{figure}

As shown in Fig.~\ref{fig:Phi05N2}, for $\phi_{\rm cl}(0)=5$ and
$\pi_{\rm cl}(0)=0$, all observables relax to equilibration values on a
timescale $t\sim10\,m^{-1}$. Throughout this work, time is measured in units
of $m^{-1}$, and we choose $a=0.25$ such that the ultraviolet lattice momentum
$\pi/a$ is well separated from the mass scale $m$. During the evolution, the
field expectation value relaxes to zero, while the pressure increases from
$-37.78$ to $19.20$ and the occupation numbers increase from $f_{p_0}=6.5$ to
$17.03$ and from $f_{p_1}=0$ to $3.3\times10^{-3}$. Notably, equilibration
proceeds sequentially in momentum space: the low-momentum $p_0$ mode
equilibrates earlier than the higher-momentum $p_1$ mode. This
sequential equilibration is quantitatively similar to the momentum-dependent
thermalization observed in kinetic theory~\cite{Kurkela:2014tea,Cabodevila:2023htm}. Since
$f_{p_1}$ is much smaller, achieving comparable precision for this observable
requires substantially more measurements and therefore sets the shot
requirement for the quantum simulations.

Although the many-body state remains pure throughout the evolution, we find that the
late-time values of all observables are accurately described by the effective
mixed state defined by the diagonal ensemble,
\begin{align}
\label{eq:rho_eq}
\rho_{\rm eq}
=
\sum_n
|c_n|^2
|E_n\rangle\langle E_n|\,,
\end{align}
where $|E_n\rangle$ is the energy eigenstate with eigenvalue $E_n$, and
$c_n=\langle E_n|\psi(0)\rangle$. The emergence of $\rho_{\rm eq}$ results from
quantum dephasing: under unitary evolution, the relative phases between
different energy eigenstates suppress the off-diagonal contributions to
expectation values through destructive interference. Consequently,
observables dynamically relax to the values predicted by the diagonal ensemble,
even though the quantum state itself remains pure. Particle production is
therefore accompanied by the transfer of energy from the coherent mean field
into quantum excitations that populate additional momentum modes.

The observed relaxation is {\it intrinsically quantum} and {\it interaction-driven}.
Coherent states reproduce the initial expectation values $\phi_{\rm cl}(0)$ and
$\pi_{\rm cl}(0)$ of the corresponding classical field configuration while
containing the minimum quantum fluctuations allowed by the uncertainty
principle. Since the classical evolution remains strictly periodic, the
relaxation of $\phi_{\rm cl}(t)$ and the associated particle production arise
from quantum many-body dynamics, including the buildup of correlations and
dephasing. In the noninteracting limit $\lambda=0$, a coherent state remains
coherent and follows the classical periodic solution, so neither relaxation
nor particle production occurs.

Equilibration of isolated many-body quantum systems and the emergence of the
diagonal ensemble have been extensively studied~\cite{Reimann:2008,Short:2011pvc,Rigol:2008,Polkovnikov:2011}.
In these approaches, equilibration is typically characterized through
long-time averages~\cite{Gogolin:2016}. In contrast, {\it no explicit time
averaging} is required here: the observables dynamically relax to the values
predicted by the diagonal ensemble. We further verify that these values agree
with the corresponding long-time averages, providing a nontrivial consistency
check. Our results extend these concepts to relativistic quantum field theory
with highly occupied classical-field initial conditions, providing, to our
knowledge, the first first-principles quantum simulation of equilibration and
particle production in this setting.

\begin{figure}[ht]
\includegraphics[width=0.46\textwidth]{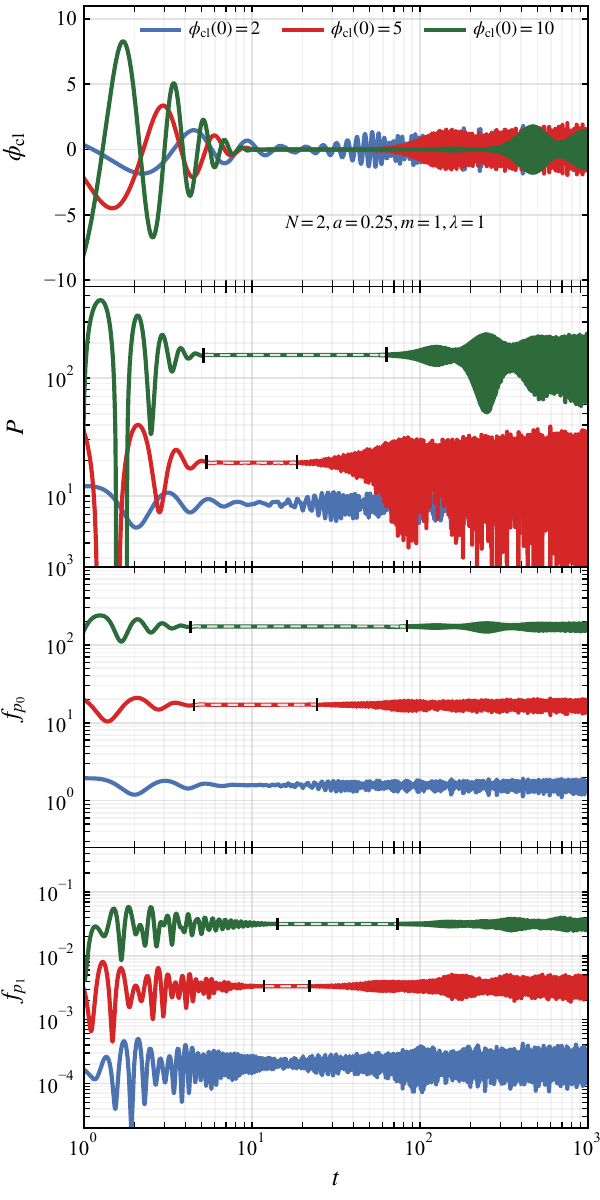}
\caption{Equilibration and quantum recurrences for $\phi_{\rm cl}(0)=2$, $5$, and $10$ with $\pi_{\rm cl}(0)=0$ on a two-site one-dimensional lattice. The parameters are $a=0.25$, $m=1$, and $\lambda=1$. Direct exponentiation of Eq.~\eqref{eq:H_HO}, involving matrices up to $2^{16}\times2^{16}$, provides an exact benchmark for the gate-based quantum algorithm. Gray dashed lines show the equilibration intervals of $P$ and $f_{\mathbf p}$, with short vertical lines marking their endpoints for $\phi_{\rm cl}(0)=5$ and $10$.}
\label{fig:resultsNx2}
\end{figure}

We next investigate the dependence of the relaxation dynamics on the initial
coherent state. For the small systems considered here, we use the exact
benchmark obtained by direct exponentiation, which is computationally more
efficient than explicit quantum-circuit simulation. As shown in
Fig.~\ref{fig:resultsNx2}, for $\phi_{\rm cl}(0)=2$, $5$, and $10$, all
observables relax toward the values predicted by the diagonal ensemble,
demonstrating the robustness of the equilibration mechanism across a broad
range of overoccupied initial states. Residual oscillations remain visible
for smaller initial field amplitudes, as exemplified by
$\phi_{\rm cl}(0)=2$, whereas increasing the amplitude suppresses these
oscillations and leads to a progressively longer-lived equilibration stage.

To quantify the duration of the equilibration stage, we define the system to
be equilibrated when, over an extended interval longer than the corresponding
classical oscillation period, the relative deviation of every observable from
its diagonal-ensemble value remains below $1\%$. According to this criterion,
the equilibration interval extends from $t=11.8\,m^{-1}$ to
$18.5\,m^{-1}$ for $\phi_{\rm cl}(0)=5$, and from
$t=14.2\,m^{-1}$ to $63.2\,m^{-1}$ for
$\phi_{\rm cl}(0)=10$. In both cases, the higher-momentum occupation number
$f_{p_1}$ is the last observable to equilibrate, following the relaxation of
$f_{p_0}$ and the pressure.

At sufficiently late times, all observables exhibit renewed oscillations,
consistent with the onset of quantum Poincar\'e recurrences in finite quantum
systems~\cite{Bocchieri:1957}. We explicitly identify such a recurrence for
$\phi_{\rm cl}(0)=2$, where the first recurrence, defined by
$|\phi_{\rm cl}(t)/\phi_{\rm cl}(0)-1|<10\%$, occurs at
$t\simeq411.6\,m^{-1}$. In contrast, no recurrence is observed yet for
$\phi_{\rm cl}(0)=5$ or $10$ within the simulated time window,
$t\leq10^3\,m^{-1}$. These revivals originate from the discrete energy
spectrum and the effectively finite-dimensional Hilbert space of the lattice
theory, which contains at most a few hundred elementary particles.
Thus, although the many-body state evolves unitarily and remains pure,
observables can remain equilibrated over long intermediate times before
finite-size coherence partially restores memory of the initial state. Our
results therefore provide an explicit demonstration of the coexistence of
equilibration and quantum Poincar\'e recurrences in small QFT systems
initialized in highly occupied coherent states.

\begin{figure}[ht]
\includegraphics[width=0.48\textwidth]{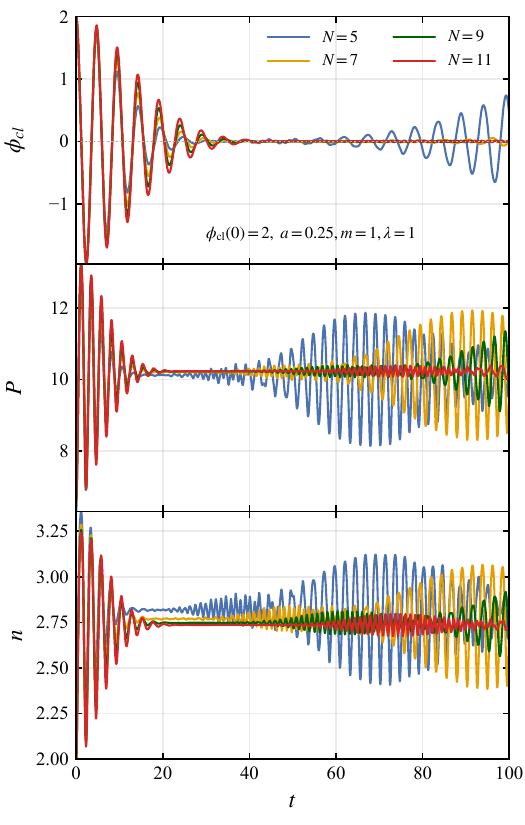}
\caption{
Equilibration with increasing system size. Shown are the field expectation value $\phi_{\rm cl}(t)$ (top), pressure $P$ (middle), and particle number density $n$ (bottom) for one-dimensional lattices with $N=5$, $7$, $9$, and $11$ sites, starting from a coherent state with $\phi_{\rm cl}(0)=2$ and $\pi_{\rm cl}(0)=0$. The parameters are $a=0.25$, $m=1$, and $\lambda=1$. The observables exhibit longer-lived equilibration stages, followed by a further delayed renewal of oscillations, as the system size increases. Tensor-network simulations use a maximal bond dimension $\chi_{\mathrm{max}}=25$ and time step $\delta t=0.01$.
}
\label{fig:resultsNlarge}
\end{figure}

{\it Quantum equilibration in larger systems.—}
For the two-site lattice, the observables corresponding to the less highly
occupied initial state, $\phi_{\rm cl}(0)=2$ and $\pi_{\rm cl}(0)=0$, exhibit
visible oscillations after the initial relaxation. This raises the question
of whether these residual oscillations arise from the smaller initial
occupation or from finite-size effects associated with the discrete energy
spectrum. To investigate this question and explore the emergence of
equilibration in larger systems, we consider one-dimensional lattices with
$N=5$, $7$, $9$, and $11$ spatial sites. For $N=5$, the dynamics can still be
obtained by direct exponentiation of the finite-dimensional Hamiltonian. For
larger lattices, direct exponentiation becomes computationally prohibitive,
and we therefore employ tensor-network (TN) simulations~\cite{Orus:2014}
using the \texttt{ITensor} library~\cite{itensor}. The $N=5$ TN results agree
well with those obtained from direct exponentiation for all observables
considered, as shown in the Supplemental Material, providing a benchmark for
the accuracy of the TN approximation used for the larger systems discussed
below.

Figure~\ref{fig:resultsNlarge} shows the time evolution of the field expectation value $\phi_{\rm cl}(t)$, the pressure $P$, and the particle number density,
\begin{align}
n
=
\frac{1}{(Na)^d}
\sum_{\mathbf p}
\langle\psi(t)|
\hat a^\dagger_{\mathbf p}
\hat a_{\mathbf p}
|\psi(t)\rangle\,,
\end{align}
for the initial coherent state with $\phi_{\rm cl}(0)=2$, $\pi_{\rm cl}(0)=0$, $a=0.25$, $m=1$ and $\lambda=1$. The qualitative evolution closely resembles that observed for the two-site lattice. The field expectation value rapidly loses amplitude, while the pressure and particle number density relax toward equilibration values. As the system size increases, the residual oscillations become progressively weaker, and the equilibration stage persists for longer times. Correspondingly, recurrence effects are shifted
to later times, consistent with the rapid growth of recurrence times with the Hilbert-space dimension. At the same time, the equilibration values of all observables exhibit clear convergence with increasing system size, indicating that finite-size effects become progressively less important.

{\it Conclusion and outlook.—}
We have presented a first-principles study of particle production from highly occupied classical fields using methods directly compatible with gate-based quantum computation. Focusing on lattice $\lambda\phi^4$ theory, we showed that particle production is accompanied by the equilibration of observables, including the field expectation value, occupation-number distribution, and pressure. Although the underlying quantum evolution remains unitary and eventually exhibits quantum recurrences, it gives rise to an equilibration stage that persists for a timescale several times longer than the initial equilibration timescale for highly occupied initial states.

Our results demonstrate that equilibration can emerge in finite isolated quantum systems with only a small number of elementary particles. The equilibration dynamics is intrinsically quantum and cannot be described by classical field theory. The relaxation of the pressure to a well-defined equilibration value suggests that an effective equation of state can emerge even in such small systems, providing insight into the active search for the smallest possible quark--gluon plasma droplets at hadron colliders~\cite{Nagle:2018nvi}. Unlike approaches based on long-time averages~\cite{Gogolin:2016}, we observe dynamical relaxation of observables within finite time windows, a feature particularly relevant for small collision systems whose lifetimes may be only moderately longer than microscopic timescales.

Several questions remain open, including how this equilibration mechanism
evolves toward larger volumes and the continuum limit, whether the observables
approach the values predicted by the Gibbs ensemble~\cite{Cuntin:2024mgm}, and
how it relates to the eigenstate thermalization
hypothesis~\cite{Deutsch:1991msp,Srednicki:1994mfb,DAlessio:2016eth}.
Quantum simulation thus opens new avenues of research toward first-principles
studies of such fundamental questions~\cite{Gogolin:2016}.

{\it Acknowledgements—}
This work is supported by the European Research Council under project ERC-2018-ADG-835105 YoctoLHC; by Maria de Maeztu excellence unit grant CEX2023-001318-M and project PID2023-152762NB-I00 funded by MICIU/AEI/10.13039/501100011033; and by ERDF/EU. It has received funding from Xunta de Galicia (CIGUS Network of Research Centres). B.W. acknowledges the support of the Ram\'{o}n y Cajal program with the Grant No. RYC2021-032271-I and the support of Xunta de Galicia under the ED431F 2023/10 project. I. C.
acknowledges the support of the Axudas de apoio á etapa predoutoral program (Ref. ED481A-2024-075). W. Q. acknowledges the support of the Marie
Sklodowska-Curie Actions under Grant No. 101109293.

\bibliography{phi.bib}

\onecolumngrid

\clearpage
\newpage
\renewcommand{\theequation}{S\arabic{equation}}
\maketitle
\begin{center}
\textbf{\large Simulating particle production from classical fields using quantum computation} \\ 
\vspace{0.05in}
{ \it \large Supplemental Material}\\ 
\vspace{0.05in}
{Iv\'an Cunt\'in, Wenyang Qian, and Bin Wu}
\end{center}

\setcounter{equation}{0}
\setcounter{figure}{0}
\setcounter{table}{0}
\setcounter{section}{0}
\setcounter{footnote}{0}
\setcounter{page}{1}


In this appendix, we provide a detailed description of the quantum simulation for the $\phi^4$ field theory including qubit encoding, initial state preparation, and resource analysis. The qubit encoding is analogous to previous treatments in Refs.~\cite{Klco:2018zqz, Bauer:2021gup, Hardy:2024ric}.

\subsection{1. Digital quantum simulation using field operator basis}

The real-time evolution for the $\phi^4$ theory can be implemented on the quantum circuit in the trotterized form by knowing the Pauli term decomposition for the field and conjugate field operators. Here, we demonstrate the circuit construction in 1+1 dimensions. 

On the qubits, the discretized Hamiltonian is mapped to a spatial lattice of $N$ sites with $n_Q$ qubits per lattice site. The field operator $\hat{\phi}_x$ at each lattice position $x$ can be constructed via
\begin{align} \label{eq:qc_phi}
    \hat{\phi}_x = -\Delta_\phi\sum_{j=0}^{n_Q-1} 2^{j-1} \sigma_j^z\,,
\end{align}
where $\sigma^z_j$ is the Pauli-Z matrix at qubit location $j$. Its time-evolved unitary becomes
\begin{align} \label{eq:qc_exp_phi}
    e^{-i\hat{\phi}_x t} = \prod_{j=0}^{n_Q-1} e^{-i\left(-2^{j-1}\Delta_\phi\,  \sigma_j^z\right) t} \equiv \prod_{j=0}^{n_Q-1} R^{Z}_{j}(-2^j t\Delta_\phi )\,,
\end{align}
where $R^{Z}_{j}(\theta)=e^{-i\sigma_j^z \theta/2}$ is the standard rotational gate.
The Pauli operator for $\hat{\phi}_x^2$ is
\begin{align} \label{eq:qc_phi2}
    \hat{\phi}_x^2 = \Delta_\phi^2 \sum_{j<k}^{n_Q-1} 2^{j+k-1} \sigma_j^z \sigma_k^z\, +C_2\,,
\end{align}
with a constant term $C_2=\Delta_\phi^2  \sum_j 4^{j-1}$. Its time-evolved circuit (up to global phase $e^{-iC_2 t}$) becomes
\begin{align} \label{eq:qc_exp_phi2}
    e^{-i\hat{\phi}^2_x t} = \prod_{j<k}^{n_Q-1}e^{-i \left(\Delta_\phi^2\sum_{j<k} 2^{j+k-1} \sigma_j^z \sigma_k^z\right)t} \equiv \prod_{j<k}^{n_Q-1} R^{ZZ}_{j,k}(\theta) (2^{j+k}t\Delta_\phi^2 ) \,,
\end{align}
where $R^{ZZ}_{j,k}(\theta)=e^{-i\sigma_j^z\sigma_k^z \theta/2}$ is the two-qubit rotation gate on qubit $j$ and $k$. 

Proceeding analogously, the $\hat{\phi}_x\hat{\phi}_{x'}$ term becomes
\begin{align} \label{eq:qc_exp_phiphi}
    e^{-it\hat{\phi}_x\hat{\phi}_{x'}} = 
    \prod_{k=0}^{n_Q-1} {
        \prod_{l=0}^{n_Q-1} {
            e^{ 
                -i t 2^{ k+l-2 }\Delta_{\varphi}^2 \sigma_{k,x}^z \sigma_{l,x'}^z
            }
        }
    }\,.
\end{align}
The Pauli operator for $\hat{\phi}^4_x$ is decomposed into three parts: the constant term, the two-body corrections, and the four-body interactions. Specifically, 
\begin{align} \label{eq:qc_phi4}
\hat{\phi}^4_x = \frac{\Delta_\phi^4}{16} \left[ \sum_{j<k<l<m} 24 \cdot 2^{j+k+l+m} \sigma_j^z \sigma_k^z \sigma_l^z \sigma_m^z + \sum_{j<k} \alpha_{jk} \sigma_j^z \sigma_k^z \right] + C_4\,,
\end{align}
where the two-body coefficient and the constant $C_4$ terms respectively are
\begin{align} 
\alpha_{jk} = 4(2^{3j+k} + 2^{j+3k}) + 12 \sum_{l \neq j,k} 2^{2l+j+k}, \quad
C_4 = \frac{\Delta_\phi^4}{16} \left[ \sum_{j=0}^{n_Q-1} 16^j + 6 \sum_{j<l}^{n_Q-1} 4^{j+l} \right].
\end{align}
The time-evolution operator for the $\hat{\phi}^4_x$ term (omitting the global phase $e^{-iC_4t}$) is
\begin{align} \label{eq:qc_exp_phi4}
e^{-i\hat{\phi}^4_x t} = \prod_{j<k<l<m} R^{ZZZZ}_{j,k,l,m}\left( 3 \cdot 2^{j+k+l+m-1} \Delta_\phi^4 t \right) \prod_{j<k} R^{ZZ}_{j,k}\left( \theta_{jk} \right)\,,
\end{align}
where the $ZZ$ rotation angle is defined by
\begin{align}\theta_{jk} = \frac{\Delta_\phi^4 t}{8} \left[ 4(2^{3j+k} + 2^{j+3k}) + 12 \sum_{l \neq j,k} 2^{2l+j+k} \right]\,,
\end{align}
and the four-qubit rotation gate $R^{ZZZZ}_{j,k,l,m}(\theta) = e^{-i \sigma_j^z \sigma_k^z \sigma_l^z \sigma_m^z \theta/2}$ is implemented via a CX-gate staircase that computes the parity of the involved qubits:
\begin{align} \label{eq:qc_exp_phi}
R^{ZZZZ}_{j,k,l,m}(\theta) = \text{CX}_{j,k} \text{CX}_{k,l} \text{CX}_{l,m} R^Z_m(\theta) \text{CX}_{l,m} \text{CX}_{k,l} \text{CX}_{j,k}\,.
\end{align}
Notably, these unitary gates involving powers of $\hat{\phi}$ operators and their combinations are exact, as all these terms commute with each other. An exemplary quantum circuit is provided in Fig.~\ref{fig:phi_circuits}.

Next, to implement the conjugate momentum operator $\hat{\pi}_x$, we use the symmetric quantum Fourier transform $\mathcal{F}_x$ at lattice site $x$ such that
\begin{align} \label{eq:qc_sqft}
    \mathcal{F}_x = e^{ -i \frac{ n_{\phi} \delta^2 }{ 2\pi } }
    \prod^{n_Q-1}_{k=0} {
        R^Z_{k,x}( -2^{k} \delta )
    } \; \mathrm{qFT}_x
    \prod^{n_Q-1}_{k=0} {
        R^Z_{k,x}( -2^{k} \delta )
    }\,,
\end{align}
where $\delta = \pi \frac{ n_{\phi} - 1 }{ n_{\phi} }$ and $n_{\phi}=2^{n_Q}$. Whereas the summation index of the standard quantum Fourier transform (qFT) runs from $0$ to $n_{\phi}-1$, that of $\mathcal{F}_x$ runs from $-(n_{\phi}-1)/2$ to $(n_{\phi}-1)/2$, matching our encoding of the field operator. It follows that any power $p$ of $\hat{\pi}_x$ can be computed via
\begin{align} \label{eq:qc_pi}
\hat{\pi}_x^p &= \mu^p \mathcal{F}_x \hat{\phi}^p_x \mathcal{F}_x^{-1}\,,
\end{align}
and consequently its time evolution unitary $\exp(-i\hat{\pi}_x^p t)$ is computed by
\begin{align} \label{eq:qc_exp_pi}
    \exp(-i\hat{\pi}_x^p t) &= \exp(-i\mu^p \mathcal{F}_x \hat{\phi}^p_x \mathcal{F}_x^{-1} t)
    =\mathcal{F}_x \exp(-i \mu^p \hat{\phi}^p_x  t) \mathcal{F}_x^{-1}\,,
\end{align}
where $\mu =(2\pi)/(a\Delta_\phi^2 n_{\phi})$ is the prefactor. Knowing the quantum circuit for both $\hat{\phi}$ and $\hat{\pi}$ operators and their powers, the real-time evolution can be constructed in a trotterized form. An exemplary quantum circuit is provided in Fig.~\ref{fig:pi_circuits}.

For the coherent initial state $\ket{\alpha}$ considered in this work, we can use the displacement operator $D(\alpha)$ acting on the vacuum (ground state) $\ket{\Omega}$ of the free Hamiltonian such that $\ket{\alpha} \equiv D(\alpha)\ket{\Omega}$. The vacuum state can be obtained using either a variational quantum eigensolver (VQE)~\cite{Peruzzo:2013bzg} or more sophisticated algorithms such as the Kitaev-Webb algorithm that builds from Gaussian states~\cite{Kitaev:2008vci}. In terms of $\phi$ and $\pi$, the displacement operator is defined as
\begin{align} \label{eq:DDisplacementOperator}
    D(\alpha) = e^{ \alpha_k \hat{a}_k^{\dagger} - \alpha_k^* \hat{a}_k } =
\mathrm{exp} \left(
    i \frac{a}{\sqrt{Na}} \sum_{x=0}^{N-1}{ 
        \cos{(pax)} \left[
            \mathrm{Im}(\alpha)\sqrt{2\omega_p}\hat{\phi}_x - \mathrm{Re}(\alpha)\sqrt{\frac{2}{\omega_p}}\hat{\pi}_x
        \right]
    }
\right).
\end{align}
For the case where $ \alpha_{\mathbf{p}} = \alpha\,\delta_{\mathbf{p},\mathbf{0}}$, and using the Baker–Campbell–Hausdorff formula 
\begin{align}
    e^Ae^B =e^{A+B+\frac12[A,B]+...},
\end{align}
Eq.~(\ref{eq:DDisplacementOperator}) reduces to 
\begin{align} \label{eq:DisplacementOperator}
D(\alpha) =
\prod_{x=0}^{N-1} { 
    \mathrm{exp} \left( -i \vartheta_{\phi}\hat{\phi}_x \right)
    \mathrm{exp} \left( -i \vartheta_{\pi}\hat{\pi}_x \right)
}
\,,
\end{align}
where $\vartheta_{\phi} = - \sqrt{\frac{ 2am }{ N} } \mathrm{Im}(\alpha)$ and $ \vartheta_{\pi} = \sqrt{ \frac{ 2a }{ Nm} } \mathrm{Re}(\alpha)$. For the finite-dimensional truncated operators used in the quantum simulation, the canonical commutation relation is not satisfied exactly. Nevertheless, the contribution from the commutator terms in the Baker--Campbell--Hausdorff expansion is found to be numerically negligible for the parameters considered, and is therefore omitted. An exemplary quantum circuit is provided in Fig.~\ref{fig:D_circuit}.

Finally, the full unitary evolution over time $t$ can be constructed in a trotterized form using $N_s$ total steps and $\delta t = t/N_s$ duration for each step,
\begin{align}
    \ket{\psi(t)} &= \exp(-iHt) \ket{\alpha} = \prod_{k=0}^{N_s-1}\exp(-iH\delta t) \ket{\alpha}\\
    &= \prod_{k=0}^{N_s-1} \Bigg\{
       \prod_{x=0}^{N-1} \left[\mathcal{F}_x \exp\left(-i\,\frac{\mu^2 \hat{\phi}_x^{2}}{2}\,a \delta t \right)\mathcal{F}_x^{-1}\right]  \exp\left[-i\sum_{x=0}^{N-1}\left( \left(\frac{m^2}{2}+1\right)\hat{\phi}_{x}^{2}+ \frac{\lambda \hat{\phi}_x^4 }{4!} - \hat{\phi}_{x}\hat{\phi}_{x+1} \right)a\delta t\right]
       \Bigg\} \ket{\alpha}.
\end{align}
In Fig.~\ref{fig:full_trotter}, we provide a schematic for the quantum circuit of the real-time evolution.

\begin{figure}[htp]
    \centering
        \[
        \vcenter{\Qcircuit @C=1.0em @R=1.0em @!R { \\
            \nghost{{q}_{0}:} & \lstick{{q}_{0}:} & \multigate{1}{e^{-it\hat{\phi}_x^2}} & \qw & \qw\\
            \nghost{{q}_{1}:} & \lstick{{q}_{1}:} & \ghost{\mathrm{e}^{-it\hat{\phi}_x^2}}        & \qw & \qw\\
        \\ }}
        \quad = \quad
        \vcenter{\Qcircuit @C=1.0em @R=0.2em @!R { \\
            \nghost{{q}_{0}:} & \lstick{{q}_{0}:} & \targ      & \gate{R^Z(2t\nu_{10})} & \targ      & \qw & \qw\\
            \nghost{{q}_{1}:} & \lstick{{q}_{1}:} & \ctrl{-1}  & \qw                                  & \ctrl{-1}  & \qw & \qw\\
        \\ }}
        \]
        \[
        \vcenter{\scalebox{0.6}{\Qcircuit @C=1.0em @R=0.2em @!R { \\
	 	\nghost{{q}_{0} :  } & \lstick{{q}_{0} :  } & \qw & \qw & \targ & \gate{R^Z({2t\rho_{3210}})} & \targ & \qw & \qw \barrier[0em]{3} & \qw & \targ & \gate{R^Z({2t\eta_{10}})} & \targ \barrier[0em]{3} & \qw & \targ & \gate{R^Z({2t\eta_{20}})} & \targ \barrier[0em]{3} & \qw & \qw & \qw & \qw \barrier[0em]{3} & \qw & \targ & \gate{R^Z({2t\eta_{30}})} & \targ \barrier[0em]{3} & \qw & \qw & \qw & \qw \barrier[0em]{3} & \qw & \qw & \qw & \qw \barrier[0em]{3} & \qw & \qw & \qw\\
	 	\nghost{{q}_{1} :  } & \lstick{{q}_{1} :  } & \qw & \targ & \ctrl{-1} & \qw & \ctrl{-1} & \targ & \qw & \qw & \ctrl{-1} & \qw & \ctrl{-1} & \qw & \qw & \qw & \qw & \qw & \targ & \gate{R^Z({2t\eta_{21}})} & \targ & \qw & \qw & \qw & \qw & \qw & \targ & \gate{R^Z({2t\eta_{31}})} & \targ & \qw & \qw & \qw & \qw & \qw & \qw & \qw\\
	 	\nghost{{q}_{2} :  } & \lstick{{q}_{2} :  } & \targ & \ctrl{-1} & \qw & \qw & \qw & \ctrl{-1} & \targ & \qw & \qw & \qw & \qw & \qw & \ctrl{-2} & \qw & \ctrl{-2} & \qw & \ctrl{-1} & \qw & \ctrl{-1} & \qw & \qw & \qw & \qw & \qw & \qw & \qw & \qw & \qw & \targ & \gate{R^Z({2t\eta_{32}})} & \targ & \qw & \qw & \qw\\
	 	\nghost{{q}_{3} :  } & \lstick{{q}_{3} :  } & \ctrl{-1} & \qw & \qw & \qw & \qw & \qw & \ctrl{-1} & \qw & \qw & \qw & \qw & \qw & \qw & \qw & \qw & \qw & \qw & \qw & \qw & \qw & \ctrl{-3} & \qw & \ctrl{-3} & \qw & \ctrl{-2} & \qw & \ctrl{-2} & \qw & \ctrl{-1} & \qw & \ctrl{-1} & \qw & \qw & \qw\\
\\ }}}
        \]
        \caption{Field operator circuits for $e^{-it\hat{\phi}^2_x}$ in Eq.~\eqref{eq:qc_exp_phi2} and $e^{-it\hat{\phi}^4_x}$ in Eq.~\eqref{eq:qc_exp_phi4} for $n_Q=2$ and $n_Q=4$ respectively, with $\nu,\eta$ and $\rho$ the exponent factors in the respective equations.
        }
        \label{fig:phi_circuits}
\end{figure}
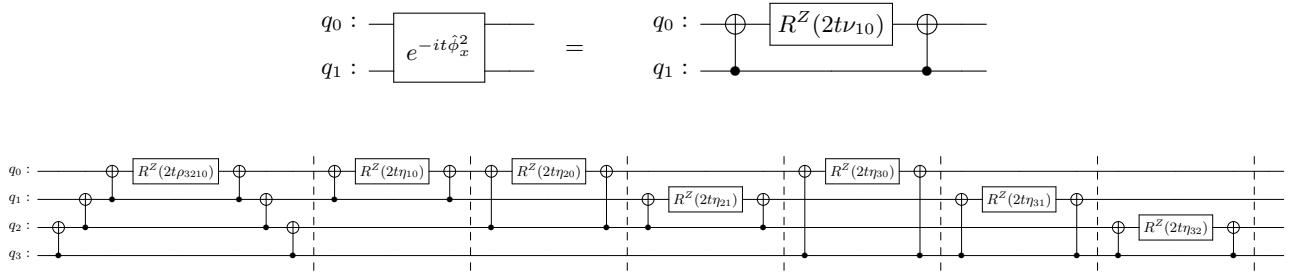
\begin{figure}
        \centering
        \[
        \vcenter{\Qcircuit @C=1.0em @R=1.0em @!R { \\
            \nghost{{q}_{0}:} & \lstick{{q}_{0}:} & \multigate{1}{e^{-it\hat{\pi}_x^2}} & \qw & \qw\\
            \nghost{{q}_{1}:} & \lstick{{q}_{1}:} & \ghost{e^{-it\hat{\pi}_x^2}}        & \qw & \qw\\
        \\ }}
        \quad = \quad
        \vcenter{\Qcircuit @C=1.0em @R=0.2em @!R { \\
            \nghost{{q}_{0}:} & \lstick{{q}_{0}:} & \multigate{1}{\mathcal{F}_x^{-1}} & \multigate{1}{\mathrm{e}^{-it\mu^2\hat{\phi}^2_x}} & \multigate{1}{\mathcal{F}_x} & \qw & \qw\\
            \nghost{{q}_{1}:} & \lstick{{q}_{1}:} & \ghost{\mathcal{F}_x^{-1}}       & \ghost{e^{-it\mu^2\hat{\phi}_x^2}} & \ghost{\mathcal{F}_x}        & \qw & \qw\\
        \\ }}
        \] 
        \[
        \vcenter{\Qcircuit @C=1.0em @R=1.0em @!R { \\
            \nghost{{q}_{0}:} & \lstick{{q}_{0}:} & \multigate{1}{\mathcal{F}_x} & \qw & \qw\\
            \nghost{{q}_{1}:} & \lstick{{q}_{1}:} & \ghost{\mathcal{F}_x}        & \qw & \qw\\
        \\ }}
        \quad = \quad
        \vcenter{\Qcircuit @C=1.0em @R=0.2em @!R { \\
            \nghost{{q}_{0}:} & \lstick{{q}_{0}:} & \gate{R^Z_x(-2^{0}\delta)} & \multigate{1}{\mathrm{qFT}_x} & \gate{R^Z_x(-2^{0}\delta)} & \qw & \qw\\
            \nghost{{q}_{1}:} & \lstick{{q}_{1}:} & \gate{R^Z_x(-2^{1}\delta)} & \ghost{\mathrm{qFT}_x}        & \gate{R^Z_x(-2^{1}\delta)} & \qw & \qw\\
        \\ }}
        \]
        \caption{Conjugate momentum operator circuits for $e^{-it\hat{\pi}_x^2}$ in Eq.~\eqref{eq:qc_exp_pi} and symmetric quantum fourier transform $\mathcal{F}_x$ in Eq.~\eqref{eq:qc_sqft} for $n_Q=2$.}
        \label{fig:pi_circuits}

        \[
    \vcenter{
    \Qcircuit @C=1.0em @R=1.0em @!R { \\
	 	\nghost{{q}_{0} :  } & \lstick{{q}_{0} :  } & \multigate{3}{D(\alpha)} & \qw & \qw\\
	 	\nghost{{q}_{1} :  } & \lstick{{q}_{1} :  } & \ghost{\mathrm{D(\alpha)}} & \qw & \qw\\
	 	\nghost{{q}_{2} :  } & \lstick{{q}_{2} :  } & \ghost{\mathrm{D(\alpha)}} & \qw & \qw\\
	 	\nghost{{q}_{3} :  } & \lstick{{q}_{3} :  } & \ghost{\mathrm{D(\alpha)}} & \qw & \qw\\
\\ }}
\qquad = \,
\vcenter{
\Qcircuit @C=1.0em @R=1.0em @!R { \\
	 	\nghost{{q}_{0} :  } & \lstick{{q}_{0} :  } & \multigate{1}{e^{-i\vartheta_{\pi}\hat{\pi}_0}} & \multigate{1}{e^{-i\vartheta_{\phi}\hat{\phi}_0}} & \qw & \qw\\
	 	\nghost{{q}_{1} :  } & \lstick{{q}_{1} :  } & \ghost{e^{-i\vartheta_{\pi}\hat{\pi}_0}} & \ghost{e^{-i\vartheta_{\phi}\hat{\phi}_0}} & \qw & \qw\\
	 	\nghost{{q}_{2} :  } & \lstick{{q}_{2} :  } & \multigate{1}{e^{-i\vartheta_{\pi}\hat{\pi}_1}} & \multigate{1}{e^{-i\vartheta_{\phi}\hat{\phi}_1}} & \qw & \qw\\
	 	\nghost{{q}_{3} :  } & \lstick{{q}_{3} :  } & \ghost{e^{-i\vartheta_{\pi}\hat{\pi}_1}} & \ghost{e^{-i\vartheta_{\phi}\hat{\phi}_1}} & \qw & \qw\\
\\ }}
\]
  \caption{Displacement operator circuit for $D(\alpha)$ in Eq.~\eqref{eq:DisplacementOperator} for $N=2$ and $n_Q=2$.}
        \label{fig:D_circuit}
\end{figure}
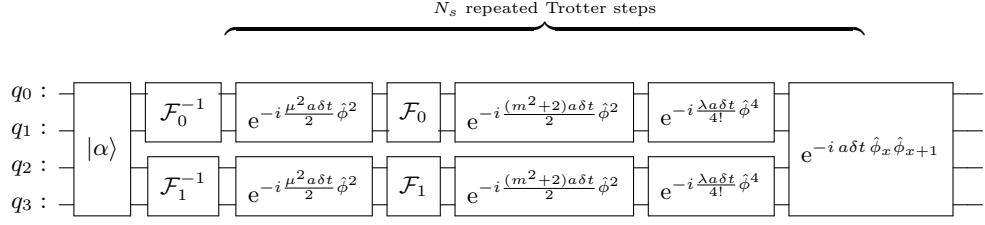
\begin{figure}
    \centering
    \[
    \hspace{5em}
    \overbrace{\hspace{26em}}^{N_s \text{ repeated Trotter steps}}
    \]
    \vspace{-1.4em}  
    \[
    \vcenter{\Qcircuit @C=0.6em @R=0.6em @!R { \\
        \nghost{{q}_{0}:} & \lstick{{q}_{0}:}
            & \multigate{3}{\ket{\alpha}}
            & \multigate{1}{\mathcal{F}_0^{-1}}
            & \multigate{1}{\mathrm{e}^{-i\frac{\mu^2 a\delta t}{2}\hat{\phi}^2}}
            & \multigate{1}{\mathcal{F}_0}
            & \multigate{1}{\mathrm{e}^{-i\frac{(m^2+2) a\delta t}{2}\hat{\phi}^2}}
            & \multigate{1}{\mathrm{e}^{-i\frac{\lambda a\delta t}{4!}\hat{\phi}^4}}
            & \multigate{3}{\mathrm{e}^{-i\,a\delta t\,\hat{\phi}_x\hat{\phi}_{x+1}}}
            & \qw & \qw \\
        \nghost{{q}_{1}:} & \lstick{{q}_{1}:}
            & \ghost{\ket{\alpha}}
            & \ghost{\mathcal{F_0}^{-1}}
            & \ghost{\mathrm{e}^{-i\frac{\mu^2 a\delta t}{2}\hat{\phi}^2}}
            & \ghost{\mathcal{F_0}}
            & \ghost{\mathrm{e}^{-i\frac{(m^2+2) a\delta t}{2}\hat{\phi}^2}}
            & \ghost{\mathrm{e}^{-i\frac{\lambda a\delta t}{4!}\hat{\phi}^4}}
            & \ghost{\mathrm{e}^{-i\,a\delta t\,\hat{\phi}_x\hat{\phi}_{x+1}}}
            & \qw & \qw \\
        \nghost{{q}_{2}:} & \lstick{{q}_{2}:}
            & \ghost{\ket{\alpha}}
            & \multigate{1}{\mathcal{F}_1^{-1}}
            & \multigate{1}{\mathrm{e}^{-i\frac{\mu^2 a\delta t}{2}\hat{\phi}^2}}
            & \multigate{1}{\mathcal{F}_1}
            & \multigate{1}{\mathrm{e}^{-i\frac{(m^2+2) a\delta t}{2}\hat{\phi}^2}}
            & \multigate{1}{\mathrm{e}^{-i\frac{\lambda a\delta t}{4!}\hat{\phi}^4}}
            & \ghost{\mathrm{e}^{-i\,a\delta t\,\hat{\phi}_x\hat{\phi}_{x+1}}}
            & \qw & \qw \\
        \nghost{{q}_{3}:} & \lstick{{q}_{3}:}
            & \ghost{\ket{\alpha}}
            & \ghost{\mathcal{F}_1^{-1}}
            & \ghost{\mathrm{e}^{-i\frac{\mu^2 a\delta t}{2}\hat{\phi}^2}}
            & \ghost{\mathcal{F}_1}
            & \ghost{\mathrm{e}^{-i\frac{(m^2+2) a\delta t}{2}\hat{\phi}^2}}
            & \ghost{\mathrm{e}^{-i\frac{\lambda a\delta t}{4!}\hat{\phi}^4}}
            & \ghost{\mathrm{e}^{-i\,a\delta t\,\hat{\phi}_x\hat{\phi}_{x+1}}}
            & \qw & \qw \\
    \\ }}
    \]
    \caption{Full real-time evolution $e^{-iHt}$ with $N=2$ and $n_Q=2$.
    }
    \label{fig:full_trotter}
\end{figure}

\subsection{2. Measurement and extraction of expectation values}

In terms of measurements, since $\sigma^z$ matrices are diagonal in the computational basis, repeated measurements of the $\sigma^z$ operator on a state $\ket{\psi} = \sum_i{c_i\ket{i}}$ yield the probability distributions
\begin{align}
    p_i = \frac{ \mathrm{counts}_{(i)} }{ \mathrm{shots} } = |c_i|^2 = |\braket{i|\psi}|^2\,,
\end{align}
where the counts$_{(i)}$ are the number of times the state $\ket \psi$ collapsed to basis state $\ket i$ and the shots are the total number of measurements performed on the circuit. 
On the other hand, the expectation value of an observable $A$ can be calculated as
\begin{align} \label{eq:EVCircuit}
    \braket{ \psi | A | \psi } = \sum_{i,j}{c^{\prime*}_jc'_i\braket{\alpha_j|A|\alpha_i} }=\sum_i {\alpha_i|c'_i|^2}\,,
\end{align}
where $\alpha_i$ and $\ket{\alpha_i}$ are the eigenvalues and eigenvectors of $A$, respectively, and $c'_i$ are the coefficients of $\ket{\psi} = \sum_i{c'_i\ket{\alpha_i}}$. We can obtain the expectation value from computational-basis sampling by applying to the circuit an operator $V$ that transforms the eigenstates of $A$ into the states of the computational basis
\begin{align}
    V\ket{\psi} = 
    \sum_i { c'_i V\ket{\alpha_i} } =
    \sum_i { c'_i \ket{i} }\,.
\end{align}
This circuit will yield the relative fractions
\begin{align}
    p_i = |\braket{i|V|\psi}|^2 =  |c'_i|^2\,,
\end{align}
which we can use to compute the expectation value of $A$ using equation (\ref{eq:EVCircuit}). In the particular case where $A$=$\sigma_q^z$, we have $V=\mathbb{I}$. Therefore, we can calculate the expectation value of $\hat{\phi}_x^2$ as
\begin{align}
    \braket{ \hat{\phi}_x^2 }_{t} = \Delta_{\varphi}^2 \sum_{q,p=0}^{n_Q-1}2^{q+p-1}\braket{ \psi(t) | \sigma_q^z\sigma_p^z | \psi(t)}\,,
\end{align}
by running a quantum circuit which implements $\ket\psi$ and computing
\begin{align}
    \braket{ \psi | \sigma_q^z\sigma_p^z | \psi } = 
    \sum_{i=0}^{n_{\phi}-1} { \lambda_i^{(q)}\lambda_i^{(p)} p_i}\,,
\end{align}
where $\lambda^{(q)}_i$ is the eigenvalue of $\sigma_q^z$ corresponding to state $\ket i$.

On the other hand, all powers of $\hat{\pi}_x$ operators can also be measured directly by prepending the quantum Fourier transform right before the measurement $(V=\mathcal{F})$. Therefore, we can calculate the expectation value of $\hat{\pi}^2_x$ as
\begin{align}
    \braket{ \hat{\pi}_x^2 }_{t} = \mu\Delta_{\varphi}^2 
    \sum_{q,p=0}^{n_Q-1}{
        2^{q+p-1}\braket{ \psi(t) | \mathcal{F}_x\sigma_q^z\sigma_p^z\mathcal{F}^{-1}_x | \psi(t)}
    },
\end{align}
by running a quantum circuit which implements $\mathcal{F}_x\ket\psi$ and computing
\begin{align}
    \braket{ \psi | \mathcal{F}_x\sigma_q^z\sigma_p^z\mathcal{F}_x^{-1} | \psi } = 
    \sum_i^{n_{\phi}-1} { \kappa_i^{(q)}\kappa_i^{(p)} p_i},
\end{align}
where $\kappa^{(q)}_i$ are the eigenvalues of $\mathcal{F}_x\sigma_q^z\mathcal{F}^{-1}_x$.

Having the ability to compute both expectations for $\braket{\hat{\phi}^p_x}$ and $\braket{\hat{\pi}^p_x}$ for any power $p$, we can calculate all other relevant physical observables $O=O_{\phi}+O_{\pi}$. 
The statistical uncertainty of an observable is estimated from the variance of its measurement outcomes. The expectation values of both $O_{\phi,\pi}$ and $O_{\phi,\pi}^2$ are obtained by evaluating the observable and its square for each measurement outcome and averaging over all shots. The statistical uncertainty is then given by~\cite{PhysRevResearch.4.033173}
\begin{align} \label{eq:ev_error}
    s(\braket{O_{\phi,\pi}})=
    \sqrt{
        \frac{ \braket{O_{\phi,\pi}^2} - \braket{O_{\phi,\pi}}^2 }{ \mathrm{shots} }
    }.
\end{align}
Since the field and conjugate-field contributions are obtained from independent quantum circuits, their statistical fluctuations are uncorrelated. Therefore, the uncertainty of observables containing both contributions, such as the energy and the particle number, is obtained by adding the corresponding variances in quadrature
\begin{align}
    s(\braket O)= 
    \sqrt{
        s^2( \braket{O_{\phi}} ) + s^2( \braket{O_{\pi}} )
    }.
\end{align}

In the ideal (noiseless) case, we use \texttt{StatevectorSampler} from \texttt{Qiskit} library~\cite{QiskitCommunity2017}, which provides access to the exact statevector of the system. By sampling from the statevector, we obtain the probabilities of all computational basis states, which are then combined with the operator’s eigenvalues to compute the expectation value.

\subsection{3. Resource estimation and scaling in digital quantum simulation}

For the real-time evolution, each trotter step simulation of $e^{-iH\delta t}$ costs $\mathcal{O}(N n_Q^4)$ CNOT gates, where $N$ is the number of lattice sites, and $n_Q$ is the number of qubits used per lattice site. The cost is dominated by the 4th power of $n_Q$ due to $\phi^4$ terms, whereas, the cost of qFT is only $\mathcal{O}(N n_Q^2)$. The total complexity for the full evolution time $t$ given a trotter error $\epsilon$ can be derived as follows. Assume a total of $r$ steps are needed. Then the error per step is $\epsilon_{step}\leq \alpha (t/r)^{p+1}$ for p-order trotterization and for some constant $\alpha$, so that the total error is $\epsilon \leq r \epsilon_{step} = \alpha t^{p+1}/r^p$. It suffices to show that the number of steps we need is $r \geq \alpha t^{p+1}/\epsilon$. Finally, the total cost of CNOT gates for an evolution time $t$ given an estimated error $\epsilon$ is
\begin{align}
    \mathrm{cost} = 
    \mathcal{O}\left(
        N n_Q^4 \frac{t^{1+1/p}}{\epsilon^{1/p}}
    \right) \,.
\end{align}

For extracting the expectation values of various observables, terms involving only the field basis scale as $\mathcal{O}(1)$, such as $\braket{\phi_x^r}$ for any integer power $r$. Observables involving the conjugate momentum basis at fixed point require one qFT, and thus the cost scales as $\mathcal{O}(n_Q^2)$. The observables such as the Hamiltonian $\braket{H}$, pressure $\braket{P}$ and $\braket{f_p}$ scale as $\mathcal{O}(Nn_Q^2)$. For all observables in this work, they scale polynomially with both $N$ and $n_Q$. Typically in measurement, the shots needed to reach accuracy of $\epsilon$ scale as $\mathcal{O}(1/\epsilon^2)$ (see Eq.~\ref{eq:ev_error}).
\begin{table}[ht]
    \centering
    \renewcommand{\arraystretch}{1.5}
    \begin{tabular}{lcc}
        \hline\hline
        Operator & CNOT gates & $R_Z$ gates \\
        \hline
        $e^{-it\hat{\phi}_x}$
            & $0$
            & $n_Q$ \\
        $e^{-it\hat{\phi}_x^2}$
            & $n_Q^2 - n_Q$
            & $\dfrac{n_Q^2 - n_Q}{2}$ \\[6pt]
        $e^{-it\hat{\phi}_x^4}$
            & $\dfrac{1}{4}n_Q^4 - \dfrac{3}{2}n_Q^3 + \dfrac{15}{4}n_Q^2 - \dfrac{5}{2}n_Q$
            & $\dfrac{n_Q(n_Q-1)(n_Q-2)(n_Q-3)}{24} + \dfrac{n_Q^2 - n_Q}{2}$ \\[6pt]
        $e^{-it\hat{\phi}_x\hat{\phi}_{x'}}$
            & $2n_Q^2$
            & $n_Q^2$ \\
        $e^{-it\hat{\pi}_x}$
            & $2n_Q^2 - 2n_Q$
            & $3n_Q^2 + 2n_Q$ \\
        $e^{-it\hat{\pi}_x^2}$
            & $3n_Q^2 - 3n_Q$
            & $\dfrac{7n_Q^2 + n_Q}{2}$ \\[6pt]
        $D(\alpha)$
            & $2n_Q^2 - 2n_Q$
            & $3n_Q^2+3n_Q$ \\
        \hline\hline
    \end{tabular}
    \caption{Gate counts for the time-evolution unitaries as a function of $n_Q$, the number of qubits per site.}
    \label{tab:gate_evolution}
\end{table}
 
\begin{table}[ht]
    \centering
    \renewcommand{\arraystretch}{1.5}
    \begin{tabular}{lcc}
        \hline\hline
        Observable & Basis rotation & CNOT gates \\
        \hline
        $\langle \hat{\phi}_x \rangle$
            & 0
            & $0$ \\
        $\langle \hat{\phi}_x^2 \rangle$
            & 0
            & $0$ \\
        $\langle \hat{\pi}_x \rangle$
            & $\mathcal{F}^{-1}$ per site
            & $n_Q^2 - n_Q$ \\
        $\langle \hat{\pi}_x^2 \rangle$
            & $\mathcal{F}^{-1}$ per site
            & $n_Q^2 - n_Q$ \\
        \hline\hline
    \end{tabular}
    \caption{Gate overhead for measuring field and conjugate-momentum observables as a function of $n_Q$, the number of qubits per site.}
    \label{tab:gate_measurement}
\end{table}

For the coherent state preparation, one could use VQE to prepare the vacuum and trotter-evolve to each $\ket{\alpha}$, which may suffice for small system size. 
Alternatively, a better approach in scaling is the following: In the first step, one prepares the vacuum state $\ket{\Omega}$ using the Kitaev-Webb algorithm~\cite{Kitaev:2008vci}. It has a known asymptotic scaling of $\mathcal{O}(N^2n_Q^2)$ for the quantum resources~\cite{Bagherimehrab:2021xlp, Jordan:2011ci}. In step two, one decomposes the coherent state:
\begin{align}
    \ket{\alpha} &= \prod_x e^{ia[\pi(0)\hat{\phi}_x - \phi(0)\hat{\pi}_x]}\ket{\Omega}=\prod_x e^{ia\pi(0)\hat{\phi}_x} e^{-ia\phi(0)\hat{\pi}_x}\ket{\Omega}=\prod_x e^{ia\pi(0)\hat{\phi}_x} e^{-ia\phi(0)\hat{\pi}_x}\ket{\Omega}\,,
\end{align}
where we have again used the fact that $e^{A+B}\approx e^A e^B$. Counting on the CNOT cost: the first term costs are exactly single-qubit rotations and thus cost nothing. The second term can be implemented with qFT and $\mathrm{qFT}^{-1}$ and thus costs $\mathcal{O}(Nn_Q^2)$ in total. Putting step 1 and step 2 together, the cost of state preparation scales as $\mathcal{O}(N^2 n_Q^2)$.

Tables~\ref{tab:gate_evolution} and~\ref{tab:gate_measurement} summarize the CNOT and $R^Z$ gate counts for each unitary block and each measurement circuit as a function of the number of qubits per site $n_Q$ and $N$ lattice sites in total.

\subsection{4. Coherent state preparation via variational quantum eigensolver}

Here, we consider the preparation of the ground state of the free (non-interacting) Hamiltonian~\cite{Li:2022ped}, before the application of a displacement operator to construct a coherent state.
Rather than approximating the ground state of the full free Hamiltonian directly, we decompose it into a local and a derivative contribution,
\begin{equation}
    H_0 = H_{\mathrm{loc}} + H_{\mathrm{der}}\,,
\end{equation}
where the local Hamiltonian is a sum of independent harmonic oscillators, whose ground state is therefore a tensor product of identical single-site ground states. Consequently, it is sufficient to determine the ground state of the one-site Hamiltonian using the VQE algorithm.

The parameterized trial state is prepared by applying a variational quantum circuit (ansatz) $U(\boldsymbol{\theta})$ to the reference state,
\begin{equation}
    |\psi(\boldsymbol{\theta})\rangle = U(\boldsymbol{\theta}) |0\rangle\,.
\end{equation}
After optimization, the resulting approximation to the one-site ground state is
\begin{equation}\label{eq:one-site-VQE}
    |\psi_{\mathrm{loc}}\rangle
    \equiv
    |\psi(\boldsymbol{\theta}^\ast)\rangle\,,
\end{equation}
and the ground state of the full local Hamiltonian is then constructed as the tensor product
\begin{equation}
    |\Psi_{\mathrm{loc}}\rangle
    =
    \bigotimes_{x=1}^{N}
    |\psi_{\mathrm{loc}}\rangle\,.
\end{equation}

The derivative term is subsequently introduced through adiabatic evolution. To this end, we define the interpolating Hamiltonian
\begin{equation}
    H(s)
    =
    H_{\mathrm{loc}}
    +
    sH_{\mathrm{der}}\,,
    \qquad
    s\in[0,1]\,,
\end{equation}
whose ground state evolves continuously from that of $H_{\mathrm{loc}}$ at $s=0$ to that of the full free Hamiltonian $H_0$ at $s=1$.

In terms of complexity, the VQE employs a linear two-local ansatz consisting of alternating layers of single-qubit $R^Y(\theta_i)=e^{-i\sigma^y\theta_i/2}$ rotations and nearest-neighbour CNOT gates, so a circuit with $L$ repetitions requires $L\left(n_Q-1\right)$ CNOT gates. See Fig.~\ref{fig:vqe_ansatz} for one possible quantum circuit used for the VQE. Following the variational preparation of the local ground state, the derivative term is switched on adiabatically. The adiabatic evolution is implemented through a digital simulation of the free Hamiltonian using a $p$th order product formula. See Fig.~\ref{fig:coherent_circuit} for the full circuit to prepare the coherent state.

\begin{figure}
\centering
\scalebox{0.8}{
\Qcircuit @C=1.0em @R=0.2em @!R { 
	 	\nghost{{q}_{0} :  } & \lstick{{q}_{0} :  } & \gate{R^Y(\theta_{1})} & \ctrl{1} & \gate{R^Y(\theta_{2})} & \qw & \ctrl{1} & \gate{R^Y(\theta_{3})} & \qw & \ctrl{1} & \gate{R^Y(\theta_{4})} & \qw & \qw & \qw & \qw & \qw & \qw\\
	 	\nghost{{q}_{1} :  } & \lstick{{q}_{1} :  } & \gate{R^Y(\theta_{5})} & \targ & \ctrl{1} & \gate{R^Y(\theta_{6})} & \targ & \ctrl{1} & \gate{R^Y(\theta_{7})} & \targ & \ctrl{1} & \gate{R^Y(\theta_{8})} & \qw & \qw & \qw & \qw & \qw\\
	 	\nghost{{q}_{2} :  } & \lstick{{q}_{2} :  } & \gate{R^Y(\theta_{9})} & \qw & \targ & \ctrl{1} & \gate{R^Y(\theta_{10})} & \targ & \ctrl{1} & \gate{R^Y(\theta_{11})} & \targ & \ctrl{1} & \gate{R^Y(\theta_{12})} & \qw & \qw & \qw & \qw\\
	 	\nghost{{q}_{3} :  } & \lstick{{q}_{3} :  } & \gate{R^Y(\theta_{13})} & \qw & \qw & \targ & \ctrl{1} & \gate{R^Y(\theta_{14})} & \targ & \ctrl{1} & \gate{R^Y(\theta_{15})} & \targ & \ctrl{1} & \gate{R^Y(\theta_{16})} & \qw & \qw & \qw\\
	 	\nghost{{q}_{4} :  } & \lstick{{q}_{4} :  } & \gate{R^Y(\theta_{17})} & \qw & \qw & \qw & \targ & \ctrl{1} & \gate{R^Y(\theta_{18})} & \targ & \ctrl{1} & \gate{R^Y(\theta_{19})} & \targ & \ctrl{1} & \gate{R^Y(\theta_{20})} & \qw & \qw\\
	 	\nghost{{q}_{5} :  } & \lstick{{q}_{5} :  } & \gate{R^Y(\theta_{21})} & \qw & \qw & \qw & \qw & \targ & \gate{R^Y(\theta_{22})} & \qw & \targ & \gate{R^Y(\theta_{23})} & \qw & \targ & \gate{R^Y(\theta_{24})} & \qw & \qw }}
\caption{Linear two-local ansatz $U(\boldsymbol{\theta})$ used in the VQE algorithm with three layers for one-site ground state in Eq.~\eqref{eq:one-site-VQE}.}
\label{fig:vqe_ansatz}
\end{figure}
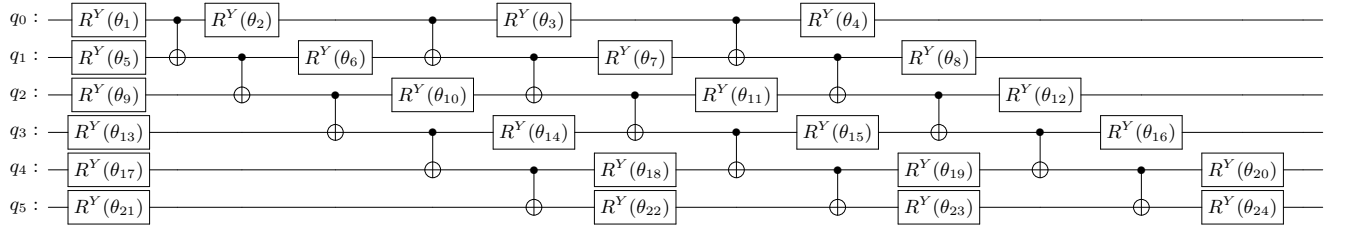
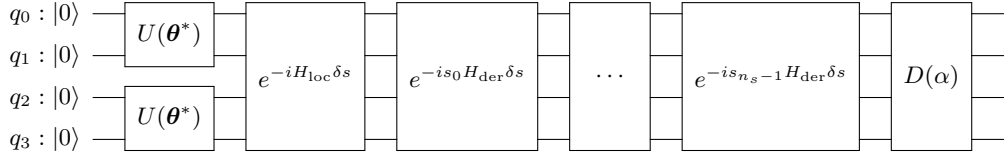
\begin{figure}
    \centering\[
\Qcircuit @C=1.3em @R=0.8em {
\lstick{q_0:\ket{0}}
& \multigate{1}{U(\boldsymbol{\theta}^\ast)}
& \multigate{3}{e^{-iH_{\rm loc}\delta s}}
& \multigate{3}{e^{-is_0H_{\rm der}\delta s}}
& \multigate{3}{\;\;\cdots\;\;}
& \multigate{3}{e^{-is_{n_s-1}H_{\rm der}\delta s}}
& \multigate{3}{D(\alpha)}
& \qw \\
\lstick{q_1:\ket{0}}
& \ghost{U(\boldsymbol{\theta}^\ast)}
& \ghost{e^{-iH_{\rm loc}\delta s}}
& \ghost{e^{-is_0H_{\rm der}\delta s}}
& \ghost{\;\;\cdots\;\;}
& \ghost{e^{-is_{n_s-1}H_{\rm der}\delta s}}
& \ghost{D(\alpha)}
& \qw \\
\lstick{q_2:\ket{0}}
& \multigate{1}{U(\boldsymbol{\theta}^\ast)}
& \ghost{e^{-iH_{\rm loc}\delta s}}
& \ghost{e^{-is_0H_{\rm der}\delta s}}
& \ghost{\;\;\cdots\;\;}
& \ghost{e^{-is_{n_s-1}H_{\rm der}\delta s}}
& \ghost{D(\alpha)}
& \qw \\
\lstick{q_3:\ket{0}}
& \ghost{U(\boldsymbol{\theta}^\ast)}
& \ghost{e^{-iH_{\rm loc}\delta s}}
& \ghost{e^{-is_0H_{\rm der}\delta s}}
& \ghost{\;\;\cdots\;\;}
& \ghost{e^{-is_{n_s-1}H_{\rm der}\delta s}}
& \ghost{D(\alpha)}
& \qw
}
\]
\caption{Preparation of the coherent state $\ket{\alpha}$ for $N=2$ and $n_Q=2$ using the optimized VQE circuit $U(\boldsymbol{\theta}^\ast)$, and for $n_s=T_{\mathrm{adiab}}/\delta s$ trotter steps, with $T_{\mathrm{adiab}}$ the total adiabatic preparation time and $s_k = k/(n_s-1)$.}
\label{fig:coherent_circuit}
\end{figure}

\subsection{5. Matrix product state simulation in harmonic oscillator basis}

To formulate the lattice $\phi^4$ theory within a finite-dimensional tensor network framework, we represent the local scalar field degree of freedom at each lattice site in a truncated harmonic oscillator (HO) basis~\cite{Klco:2018zqz,Hardy:2024ric}. This provides a controlled and systematically improvable representation of the bosonic Hilbert space. For numerical treatment, the HO is truncated to $\{\ket{0}, \ket{1}, \ket{2}, ..., \ket{N_\mathrm{max} -1}\}$ where $N_\mathrm{max}$ defines the local bosonic cutoff at each momentum mode. In principle, $N_{\max}$ need not be uniform, where different truncations can be assigned to each mode. In practice, we can exploit this by allocating larger $N_{\max}$ to the low-momentum (zero) mode where the coherent-state population is largest, and smaller $N_{\max}$ to high-momentum modes. The full Hamiltonian can be built directly using matrix product state (MPS) on a bosonic lattice using \texttt{ITensor}~\cite{itensor} library.

We use an MPS chain consisting of $N$ sites, one per momentum mode
$k = 0, 1, \ldots, N-1$, with corresponding momentum
\begin{align}
    p_k = \frac{2\pi}{Na}\left\lfloor \frac{k - (N-1)}{2} \right\rfloor\,,
\end{align}
and lattice dispersion relation $\omega_k = \sqrt{4\sin^2(p_k a/2)/a^2 + m^2}$ on the lattice. The bond indices of the MPS carry the entanglement between momentum modes and grow in dimension during time evolution up to a maximum bond dimension $\chi_\mathrm{max}$ that controls the truncation of many-body correlations.

In the HO basis, the field operator at physical site $x$ is expanded as
\begin{align}
    \hat{\phi}_{x}
    = \frac{1}{\sqrt{Na}} \sum_{k=0}^{N-1}
      \frac{e^{ip_k x}}{\sqrt{2\omega_k}}
      \bigl(\hat{a}_k + \hat{a}^\dagger_k\bigr)\,,
\end{align}
where $\hat{a}_k$ and $\hat{a}_k^\dagger$ are the annihilation and creation operators
at mode $k$, represented as $(N_{\max}\times N_{\max})$ matrices
$(\hat{a}_k)_{nm} = \sqrt{n}\,\delta_{m,n-1}$ within the truncated space.
The conjugate momentum $\hat{\pi}_{\mathbf{x}}$ is obtained analogously with the
replacement $(\hat{a}_k + \hat{a}_k^\dagger) \to -i(\hat{a}_k - \hat{a}_k^\dagger)$
and weight $\sqrt{\omega_k/2}$ instead of $1/\sqrt{2\omega_k}$.
The full Hamiltonian separates into a free part and an interaction,
\begin{align}
    H = H_0 + H_{\rm int}\,, \qquad
    H_0 = \sum_k \omega_k \Bigl(\hat{a}^\dagger_k \hat{a}_k + \tfrac{1}{2}\Bigr)\,,
    \qquad
    H_{\rm int} = \frac{\lambda}{4!} a \sum_{x} \hat{\phi}_{\mathbf{x}}^4\,.
\end{align}
$H_0$ is diagonal in the Fock basis and contributes only on-site terms to the MPO.
The quartic interaction $H_{\rm int}$, when expanded in momentum modes, generates
$\mathcal{O}(N^4)$ four-mode interaction terms of the form
$\hat{a}^{(\dagger)}_{k_1}\hat{a}^{(\dagger)}_{k_2}
 \hat{a}^{(\dagger)}_{k_3}\hat{a}^{(\dagger)}_{k_4}$
with $k_1+k_2+k_3+k_4 = 0 \pmod{N}$ (momentum conservation).
The full Hamiltonian is constructed as a matrix product operator (MPO) using the
\texttt{OpSum} interface of the \texttt{ITensor} library~\cite{itensor}, which automatically
compresses the operator into an efficient MPO representation.

The system is initialized in a coherent state with displacement concentrated on a single momentum mode. Starting from the vacuum $\ket{0}^{\otimes N}$, which is the natural ground state in the HO basis, the
displacement operator
\begin{align}
    D(\alpha_k) = e^{\,\alpha_k \hat{a}_k^\dagger - \alpha_k^* \hat{a}_k}
\end{align}
is applied independently at each site $k$, with $\alpha_k = \alpha_0\,\delta_{k,k_0}$ nonzero only at the zero-momentum mode $k_0 = \lfloor(N-1)/2\rfloor + 1$ (the central index of the momentum array, which corresponds to $p_{k_0}=0$ for the momentum ordering used here).
In practice, $D(\alpha_k)$ is exponentiated directly as a local $(N_{\max}\times N_{\max})$ dense matrix and applied to the MPS via a single-site gate, so that the resulting state is still a product MPS with bond dimension one at $t=0$. The displacement amplitude $\alpha_0 \in \mathbb{C}$ is fixed by the classical initial conditions $\phi(0)$ and $\pi(0)$ via
\begin{align}
    \mathrm{Re}(\alpha_0) = \sqrt{\frac{Nam}{2}}\,\phi(0)\,,
    \qquad
    \mathrm{Im}(\alpha_0) = \sqrt{\frac{Na}{2m}}\,\pi(0)\,,
\end{align}
so that $\langle\hat\phi_{\mathbf{x}}\rangle = \phi(0)$ and $\langle\hat\pi_{\mathbf{x}}\rangle = \pi(0)$ uniformly at every lattice site,
matching the spatially homogeneous classical initial condition. All other modes $k\neq k_0$ remain in the vacuum.

\begin{figure}[t]
\includegraphics[width=0.5\textwidth]{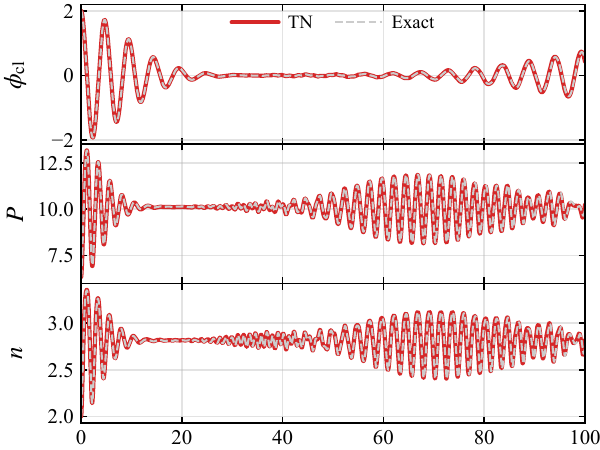}
\caption{
Comparison between the tensor-network (TN) and direct-exponentiation (exact) results for an initial coherent state with $\phi_{\rm cl}(0)=2$ and $\pi_{\rm cl}(0)=0$ on a one-dimensional lattice with $N=5$, $a=0.25$, $m=1$, and $\lambda=1$. The agreement validates the TN approach for the observables considered in the main text.
}
\label{fig:TN_vs_Exact}
\end{figure}

Real-time evolution is performed using the two-site time-dependent variational
principle (TDVP)~\cite{Haegeman:2011zz,Haegeman:2015ezw} as implemented in the \texttt{ITensorTDVP} package. At each time step $\Delta t$, the MPS is evolved under
$e^{-iH\Delta t}$ with a reverse-step correction to reduce Trotter error, and the bond dimension is allowed to grow up to the maximum $\chi_\mathrm{max}$. 
At each time step, the following expectation values are computed and recorded:
the energy E, the pressure P,
the mode occupation $\langle\hat{n}_k\rangle = \langle\hat{a}^\dagger_k\hat{a}_k\rangle$,
and the field and conjugate-momentum expectation values
$\langle\hat{\phi}_x(t)\rangle$ and $\langle\hat{\pi}_x(t)\rangle$
as MPO expectation values contracted with the current MPS.
Notably, in particular, the phase-space distribution at momentum $p_k$ is
\begin{align}
    f_{p_k}(t) = \langle\psi(t)|\hat{a}^\dagger_k \hat{a}_k|\psi(t)\rangle \equiv \langle\psi(t)|\hat{n}_k|\psi(t)\rangle\,,
\end{align}
obtained directly from the on-site number operator, which is diagonal in the HO basis and hence inexpensive to evaluate. Field expectation values $\langle\hat{\phi}_x\rangle$ involve a sum over all modes and are computed by contracting the corresponding MPO with the MPS. See Fig.~\ref{fig:TN_vs_Exact} for a comparison between the tensor-network and direct-exponentiation results on a lattice of $N=5$ sites.

\subsection{6. Analytical derivation for initial conditions for fixed energy density}
Lastly, we provide analytical derivations for initial conditions at fixed initial energy density. In our setup, the initial energy density is
\begin{align} \label{eq:EnergyDensity}
    \varepsilon_0 = \frac{\mathcal{E}_0}{Na} = 
\frac{1}{2} \left( 
    m^2\braket{\hat{\phi}^2(0)} + \braket{\hat{\pi}^2(0)} + \frac{1}{a^2}\braket{\nabla\hat{\phi}(0)} 
\right) + \frac{\lambda}{4!}\braket{\hat{\phi}^4(0)}\,.
\end{align}
For $\braket{\hat{\phi}^2}$, we have
\begin{align}
    \braket{\alpha_0|\hat{\phi}^2 |\alpha_0} 
    =\frac{1}{Na}\frac{1}{2m} (\alpha_0^2 +2|\alpha_0|^2+ (\alpha_0^*)^2) +
    \frac{1}{2Na} \sum_p { \frac{1}{\omega_p} }
    = 
    \frac{ 2 }{ Nam }  \mathrm{Re}^2(\alpha) + \Omega_{\phi} = \phi_0^2 + \Omega_{\phi}\,.
\end{align}
with $\phi_0 \equiv \phi(0)$ and
\begin{align}
    \Omega_{\phi} = \frac{1}{2Na} \sum_p { \frac{1}{\omega_p} }\,.
\end{align}
Likewise, for the $\braket{\hat{\pi}^2(0)}$ term:
\begin{align}
    \braket{\alpha_0|\hat{\pi}^2 |\alpha_0} = 
    \frac{ 2m }{ Na }  \mathrm{Im}^2(\alpha) + \Omega_{\pi} = \pi_0^2 + \Omega_{\pi}\,,    
\end{align}
with $\pi_0 \equiv \pi(0)$ and
\begin{align}
    \Omega_{\pi} = \frac{1}{2Na} \sum_p { \omega_p }\,.
\end{align}
For $\braket{\hat{\phi}^4(0)}$, we have
\begin{align}
    \braket{\alpha_0|\hat{\phi}^4 |\alpha_0} =
        \frac{1}{4N^2a^2}\left[ 
        \alpha_0^4 + 4\alpha_0^*\alpha_0^3 + 6|\alpha_0|^4 + 4(\alpha_0^*)^3\alpha_0  +
        (\alpha_0^*)^4 + 6\Omega_{\phi}(\alpha_0^2 +2|\alpha_0|^2+ (\alpha_0^*)^2) + 3\Omega_{\phi}^2
    \right]\,,
\end{align}
which simplifies to $\braket{\hat{\phi}^4(0)}=\phi_0^4 + 6\phi_0^2\Omega_{\phi} + 3\Omega_{\phi}^2$. Similarly, the derivative term yields
\begin{align}
    \hat{\phi}_{x+1} - \hat{\phi}_{x} = 
    \frac{1}{\sqrt{Na}}\sum_{p} { \sqrt{\frac{1}{2\omega_p}} 
    \left( 
        a_p e^{ip\mathrm{a}x}(e^{ipa}-1)+a^{\dagger}_{p}e^{-ip\mathrm{a}x}(e^{-ipa}-1) 
    \right) } \,.
\end{align}
Since $e^{\pm ipa}-1=0$ when $p=0$, the only non-zero term will come from the commutator obtained by replacing $\braket{a_pa^{\dagger}_{p'}}=[a^{\dagger}_{p'} ,a_p]+ a^{\dagger}_{p'} a_p$, that is
\begin{align}
    \braket{(\hat{\phi}_{x+1} - \hat{\phi}_{x})^2} &= 
    \frac{1}{2Na} \sum_p { \frac{1}{\omega_p} }
    \left[
        (e^{ipa}-1)(e^{-ipa}-1) + (e^{-ipa}-1)(e^{ipa}-1)
    \right]  \\\notag
    &=\frac{1}{Na} \sum_p { \frac{1}{\omega_p} }(e^{ipa}-1)(e^{-ipa}-1)= 
    \frac{1}{Na} \sum_p { \frac{1-\cos{(pa})}{\omega_p} }\,.
\end{align}
Therefore, we can fix our initial energy density by setting the values of $\mathcal{E}_0$ and $\phi_0$ and solving (\ref{eq:EnergyDensity}) for $\pi_0$, which gives us
\begin{align}
    \pi_0^2 &= \frac{2\mathcal{E}_0}{Na} - 2\frac{\lambda}{4!}\braket{\hat{\phi}^4(0)} -
\left( 
    m^2\braket{\hat{\phi}^2(0)} + \Omega_{\pi} + \frac{1}{a^2}\braket{\nabla\hat{\phi}(0)} 
\right) \\\notag
&= \frac{2\mathcal{E}_0}{Na} - \phi_0^2
\left(
    \frac{\lambda}{12}\phi_0^2 + \frac{\lambda}{2}\Omega_{\phi} + m^2
\right) - \Omega{\phi}
\left(
    \Omega{\phi}\frac{\lambda}{4} + m^2
\right) - \Omega_{\pi} - \frac{1}{Na^3} \sum_p { \frac{1-\cos{(pa})}{\omega_p} }\,.
\end{align}

\end{document}